\documentstyle[preprint,revtex]{aps}
\begin{document}
\baselineskip=24pt plus 3pt minus 1pt
\hfill{\bf KFKI-RMKI-25-APR-1993}
\bigskip\bigskip\bigskip
\begin{title}
Comment on "Consistent Interpretation
of Quantum Mechanics Using Quantum Trajectories"
\end{title}
\author{Lajos Di\'osi}
\begin{instit}
KFKI Research Institute for Particle and Nuclear Physics\\
H-1525 Budapest 114, POB 49, Hungary\\
electronic-mail: diosi@rmki.kfki.hu
\end{instit}
\bigskip
\pacs{PACS numbers: 03.65.Bz, 05.30.-d}
\bigskip

In a recent Letter [1], Griffiths presents a generalization
of the consistent
history approach to quantum mechanics. He defines families of
quantum trajectories using bases ${\vert\phi_j^\alpha>}$
for the quantum Hilbert
space at different times $t_j\ (j=1,2,\dots,n)$ chosen so that an
appropriate noninterference condition (NIC) is satisfied.

I think one can easily construct all possible {\it complete families}
satisfying Griffiths' NIC.

Let us start from the general identity valid for the unitary evolution
operator given by Eq.\ (1) of the Letter:
\begin{equation}
<\phi_n^{\alpha_n}\vert U(t_n-t_1)\vert\phi_1^{\alpha_1}>\ =
\sum_{\alpha_2}\sum_{\alpha_3}\dots\sum_{\alpha_{n-1}}\prod_{j=1}^{n-1}
<\phi_{j+1}^{\alpha_{j+1}}\vert U(t_{j+1}-t_j)\vert\phi_j^{\alpha_j}>,
\end{equation}
where the summation goes over all {\it paths} with fixed endnodes.
If the LHS is nonzero then if follows from the NIC of the Letter
that only a single path contributes to the sum on the RHS. This path is
a {\it quantum trajectory}.
Let us denote it by $(1,\alpha_1),(2,\alpha_2),...,(n,\alpha_n)$. Hence,
\begin{equation}
<\phi_n^{\alpha_n}\vert U(t_n-t_1)\vert\phi_1^{\alpha_1}>\ =
\prod_{j=1}^{n-1}
<\phi_{j+1}^{\alpha_{j+1}}\vert U(t_{j+1}-t_j)\vert\phi_j^{\alpha_j}>.
\end{equation}
If there were
more then one path of nonzero contribution,
connecting the same endnodes
then, according to the NIC, there would be no quantum trajectory
connecting the nodes $(1,\alpha_1)$ and $(n,\alpha_n)$. This, in turn, would
contradict to the claim of the Letter that the
sum of weights over an
{\it elementary} family [see Fig.\ 1(c) of Griffiths] is $1$.

I show that the general solution of Eq.\ (2) is surprisingly simple.
Let us introduce new basis vectors by the following unitary transformations:
$\vert\psi_j^\alpha>=U(t_1-t_j)\vert\phi_j^\alpha>$ for $j=1,2,\dots,n$.
Then Eq.\ (2) can be rewritten as
\begin{equation}
<\psi_n^{\alpha_n}\vert\psi_1^{\alpha_1}>\ =
\prod_{j=1}^{n-1}
<\psi_{j+1}^{\alpha_{j+1}}\vert\psi_j^{\alpha_j}>.
\end{equation}

Remind that, by assumption, the LHS is not zero. For the moment, forget
the upper labels $\alpha_j$, and consider the equation
\begin{equation}
<\psi_n\vert\psi_1>\ =
\prod_{j=1}^{n-1}
<\psi_{j+1}\vert\psi_j>.
\end{equation}
It is easy enough to inspect that this equation is satisfied only if
the first $m$ states $\vert\psi_1>,\dots,\vert\psi_m>$
are {\it identical} with
each other and, similarly, the remaining $n-m$ states $\vert\psi_{m+1}>,
\dots,\vert\psi_n>$ are also identical ones, where $1\le m \le n$. The
cases $m=1$ and $m=n$ are marginal: all n states $\vert\psi_j>
\ (j=1,2,\dots,n)$ will be the same if the first $(j=1)$ and the last $(j=n)$
ones have coincided.

Let us summarize our result in terms of the original bases of
Griffiths. All possible {\it complete family} of quantum trajectories
can be constructed in the following way. Choose and fix
the initial
basis $\vert\phi_1^{\alpha}>$ and the final basis $\vert\phi_n^\alpha>$.
Calculate the general matrix element
$<\phi_n^{\alpha_n}\vert U(t_n-t_1)\vert\phi_1^{\alpha_1}>$.
If it is zero then no quantum trajectory connects
the corresponding
endnodes $(1,\alpha_1)$ and $(n,\alpha_n)$. If the modulus of the matrix
element is $1$  then the following trajectory will connect the
endnodes:
\begin{equation}
\vert\phi_j^\alpha>\ =U(t_{j+1}-t_1)\vert\phi_1^\alpha>,
                                                   \ \ \ (j=1,2,\dots,n-1).
\end{equation}
If, alternatively, the modulus of the matrix element is less then
unity (but nonzero) then there is a less trivial
connective quantum trajectory:
\begin{eqnarray}
\vert\phi_j^\alpha>\ &=&U(t_j-t_1)\vert\phi_1^\alpha>
                                           \ \ \ (j=1,2,\dots,m),\nonumber\\
\vert\phi_j^\alpha>\ &=&U(t_j-t_n)\vert\phi_n^\alpha>
                                           \ \ \ (j=m+1,m+2,\dots,n),
\end{eqnarray}
with $1<m<n$. The Eqs.\ (5) and (6) yield the most general complete
families conform with the proposal of the Letter [1].
(Both altering the phases of the basis vectors (6)
and relabelling them are allowed but physically irrelevant.)

Seemingly, the quantum trajectories
of Eq.\ (5) correspond to the ordinary unitary evolution of the
state vector, cf.\  the Eq.\ (2) of the Letter.
With the initial condition $\vert\Psi(0)>\ =\vert\phi_1^{\alpha}>$,
the equations $\vert\Psi(t_j)>\ = \vert\phi_j^\alpha>$ hold for all $t_j$.
In case of Eq.\ (6) something happens to the wave function
$\vert\Psi(t)>$ in the interval $t_m<t<t_{m+1}$ otherwise the
evolution is unitary and concludes as
$\vert\Psi(t_j)>\ =\vert\phi_n^\alpha>$ at time $t_n$.

Regarding the triviality of the complete families of quantum
trajectories, there is a temptation to re-reinterpret them
on the conventional language of von Neumann measurement
theory (cf.\ Ref.\ [2]).
I am afraid
Griffiths' proposal [1] has not got farther than the ordinary
theory of quantum measurement.

This work was supported by the Hungarian Scientific Research
Fund under Grant OTKA 1822/1991.


\begin{references}
\item\ \ R. B. Griffiths, Phys. Rev. Lett. {\bf70}, 2201 (1993).\smallskip
\item\ \ L. Di\'osi, Phys. Lett.{\bf280B}, 71 (1992).
\end{references}
\end{document}